\documentclass[conference]{IEEEtran}
\IEEEoverridecommandlockouts
\usepackage{cite}
\usepackage{amsmath,amssymb,amsfonts,amsthm}
\usepackage{algorithmic}
\usepackage{graphicx}
\usepackage{textcomp}
\usepackage{xcolor}
\usepackage{todonotes}
\usepackage{calrsfs}
\usepackage[colorlinks,allcolors=black]{hyperref}

\newtheorem{theorem}{Theorem}
\newtheorem{corollary}{Corollary}
\newtheorem{proposition}{Proposition}

\theoremstyle{definition}
\newtheorem{definition}{Definition}
\newtheorem{example}{Example}
\newtheorem{notation}{Notation}
\newtheorem{remark}{Remark}

\newcommand{\numberset}{\mathbb}

\newcommand{\Z}{\numberset{Z}}

\newcommand{\F}{\numberset{F}}

\newcommand{\mC}{\mathcal{C}}

\newcommand{\mA}{\mathcal{A}}
\newcommand{\mN}{\mathcal{N}}

\newcommand{\mF}{\mathcal{F}}

\newcommand{\st}{\, : \,}

\newcommand{\mE}{\mathcal{E}}

\newcommand{\mV}{\mathcal{V}}

\newcommand{\inn}{\textnormal{in}}
\newcommand{\out}{\textnormal{out}}

\newcommand{\CA}{\textnormal{C}_1}

\newcommand{\bfT}{{\bf T}}
\newcommand{\bfx}{{\bf x}}

\newcommand{\alphabet}{\mA}

\newcommand{\card}[1]{|#1|}
\newcommand{\T}{^\intercal}
\newcommand{\sprod}[2]{{#1}\T{#2}}

\def\BibTeX{{\rm B\kern-.05em{\sc i\kern-.025em b}\kern-.08em
T\kern-.1667em\lower.7ex\hbox{E}\kern-.125emX}}

\begin{document}

\title{Network Coding: An Optimization Approach}

\author{
\IEEEauthorblockN{Christopher Hojny, Altan B. K\i l\i \c{c}, Alberto Ravagnani \thanks{A. B. K. is supported by the Dutch Research Council through grant VI.Vidi.203.045. A. R. is supported by the Dutch Research Council through grants VI.Vidi.203.045 and
OCENW.KLEIN.539,
and by the Royal Academy of Arts and Sciences of the Netherlands.}}
\IEEEauthorblockA{\textit{Department of Mathematics and Computer Science} \\
\textit{Eindhoven University of Technology, the Netherlands}
%\\
%Eindhoven, the Netherlands
\\
\texttt{\{c.hojny,\, a.b.kilic,\, a.ravagnani\}@tue.nl}}
}

\maketitle

%%%%%%%%%%%%%%%%%%%%%%%%%%%%%%%%%%%%%%%%%%%%%

\begin{abstract}
We consider the problem of
computing the capacity of a coded, multicast network
over a small alphabet. We introduce a novel approach 
to this problem based on mixed-integer programming. As an application of our approach, we recover, extend and refine
various results that were previously obtained with case-by-case analyses or specialized arguments, giving evidence of the wide applicability of our approach and its potential. We also provide two simple ideas that reduce the complexity of our method for some families of networks. We conclude the paper by outlining a research program we wish to pursue in the future to investigate the capacity of large networks affected by noise, based on the  approach proposed in this paper.
\end{abstract}

%%%%%%%%%%%%%%%%%%%%%%%%%%%%%%%%%%%%%%%%%%%%


\section{Introduction}
\label{sec:intro}

\textit{Network coding} is a communication strategy
where the intermediate nodes of a network are allowed to combine (i.e., to \textit{code}) alphabet symbols before sending them towards the receivers.
When the alphabet size is a finite field and the functions implemented by the intermediate nodes are linear over that field, the strategy is called \textit{linear} network coding; see~\cite{b8, b4, b28, b26,b2,b7} among many others and~\cite{b12} as a general reference.

It has been shown that linear network coding achieves the network's capacity, if the underlying field is large enough~\cite{b4,b8,b7}.
Moreover, an efficient algorithm for designing capacity achieving functions was given in~\cite{b26}, where the field size is required to be at least as large as the number of terminals.

It is a long-standing open problem to determine the capacity of a coded network
over small alphabets. In the last decade some progress on this problem has been made, e.g. in~\cite{b9, b10, b11, b13}, even though a systematic approach seems to be missing and currently out of reach.
This problem is relevant also for very small networks, which, as recently shown in~\cite{net_dec}, can be used as the ``building blocks'' of more general network theories.

In this paper,
we present a new and effective framework using mixed-integer programming~\cite{b29,b30} for studying networks and their capacity over small alphabet sizes. 
As an application, we recover, extend, and refine various results on the capacity of certain ``nasty'' networks, previously investigated with \textit{ad-hoc} methods.

This paper is the first stepping stone on the applications of mixed-integer programming techniques to network coding problems. An ambitious, long-term program is outlined at the end of the paper. 

The remainder of this document is organized as follows. Section~\ref{sec:network} introduce the necessary notations, the setup for our problem, and state some 
preliminary results. In Section~\ref{sec:model} we present a mixed-integer programming model for computing the capacity of communication networks. Section~\ref{sec:coding} 
contains two observations 
that reduce the running time
of our algorithm. In Section~\ref{sec:examples} we apply our model in some concrete instances and give evidence of its applicability.
Section~\ref{sec:conclusion} 
describes the future research program.

%%%%%%%%%%%%%%%%%%%%%%%%%%%%%%%%%%%%%%%%%%%%

\section{Network Coding}
\label{sec:network}
We formally define communication networks 
and their capacity. We focus on single-source networks, although most of the paper can be generalized to multiple sources.
 
\begin{definition}
\label{def:network}
A \textbf{(single-source)} \textbf{network}  is a 4-tuple $\mN=(\mV,\mE, S, \bfT)$ 
that satisfies the following properties:
\begin{enumerate}[\IEEEsetlabelwidth{7)}]
\item $(\mV,\mE)$ is a finite, directed, acyclic multigraph;
\item $S \in \mV$ is the \textbf{source};
\item $\bfT \subseteq \mV$ is the set of \textbf{terminals};
\item $|\bfT| \ge 1$ and $S \notin \bfT$;
\item the source does not have incoming edges, and terminals do not have outgoing edges;
\item \label{prnE} for any~$T \in \bfT$, there exists a directed path from~$S$ to~$T$;
\item for every~$V \in \mV \setminus (\{S\} \cup \bfT)$, there exists a directed path from~$S$ to~$V$ and from~$V$ to some~$T \in \bfT$. 
\end{enumerate}
The elements of~$\mV$ are called \textbf{vertices}, the elements of~$\mV \setminus (\{S\} \cup \bfT)$ are called \textbf{intermediate} vertices, and the elements of~$\mE$ are called \textbf{edges}.
\end{definition}

The edges of a network carry exactly one symbol 
from an \textbf{alphabet} $\mA$
(a finite set with cardinality at least 2).
We focus
on multicast networks and 
we assume them to be 
delay-free. 

\begin{notation}
Throughout this paper, $\mN=(\mV,\mE, S, \bfT)$ always denotes a network and $\mA$ a network alphabet.
\end{notation}

\begin{remark}
\label{rem:order}
A partial order on the edges of $\mN$ is given, for 
$e,e' \in \mE$,
by  $e \preccurlyeq e'$
if~$\mN$ contains a directed path that starts with~$e$ and ends with~$e'$.
By the order-extension principle~\cite{b1}, this partial order can be extended to a total order $\le$.
\end{remark}

The intermediate vertices of a network are allowed to process incoming alphabet symbols according to functions assigned to them. This strategy is called network coding, see~\cite{b2,b4,b7,b8,b12, b14,b17,b18,b19,b20, b27,b28} among many others.

\begin{definition}
\label{def:nc}
A \textbf{network code} for~$(\mN,\mA)$ is a set of functions~$\mF=\{\mF_V \st V \in \mV \setminus (\{S\} \cup \bfT)\}$, where $$\mF_V : \mA^{|\inn(V)|} \to \mA^{|\out(V)|} \quad \mbox{for all $V \in \mV \setminus (\{S\} \cup \bfT)$}.$$
Here and throughout the paper, $\inn(V)$ and~$\out(V)$ denote the set of incoming and outgoing edges of~$V$, respectively. 
\end{definition}

Note that a network code uniquely specifies how
incoming symbols are processed by a vertex, whenever a total order extending $\preccurlyeq$ has been chosen; see e.g. \cite{b20}. None of the results in this paper depends on the specific choice of the order extension. 

The following notation is necessary to formally describe the functions induced by a network code.

\begin{notation} 
\label{not:channel}
Let $T \in \bfT$ be a terminal and let~$\mF$ be a network code for~$(\mN,\mA)$. We denote by
$$\Omega[\mN, \mF, S \to T] : \mA^{|\out(S)|} \rightarrow \mA^{|\inn(T)|}$$
the function representing the transmission from~$S$ to~$T$, when the network code $\mF$ is used. In the terminology of~\cite{b20}, 
$\Omega[\mN, \mF, S \to T]$ is a \textit{deterministic channel}, and therefore identified with a function.
\end{notation}

The transmission is initialized by the source of the network, which emits symbols over its outgoing edges.

\begin{definition}
\textbf{An (outer) code} for 
$(\mN,\mA)$ 
is a non-empty subset $$\mC \subseteq \mA^{|\out(S)|}.$$ Given a network code~$\mF$ for~$(\mN,\mA)$, a pair~$(\mC,\mF)$ is \textbf{unambiguous} for~$(\mN,\mA)$ if
for all~$T \in \bfT$
we have
$$\Omega[\mN, \mF, S \to T](c) \neq \Omega[\mN, \mF, S \to T](c')$$ for all~$c,c' \in \mC$ with $c \neq c'$.
\end{definition}

We use the celebrated Butterfly Network \cite{b2} to illustrate the concepts introduced so far. 
\begin{figure}[htbp]
\centerline{\includegraphics[width=0.5\textwidth]{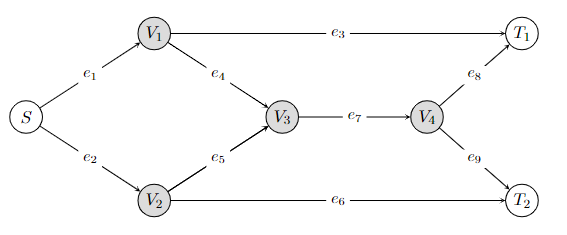}}
\caption{The Butterfly Network.}
\label{fig:butterfly}
\end{figure}

\begin{example}[The Butterfly Network]
\label{ex:butterfly}
Let~$\mN=(\mV,\mE, S, \bfT)$ be the Butterfly Network of Figure~\ref{fig:butterfly}, where the edges are ordered according to their indices. The vertex set is $\mV = \{S,V_1,V_2,V_3,V_4,T_1,T_2\}$ and the set of edges is
$\mE = \{e_1,e_2,e_3,e_4,e_5,e_6,e_7,e_8,e_9\}$. We have $\bfT$ = $\{ T_1,T_2\}$.
Let~$\mA = \F_3$ and consider the network code $$\mF=\{\mF_1,\mF_2,\mF_3,\mF_4\},$$ where~$\mF_1(a) = (2a,a)$, $\mF_2(a) = (a,2a)$, $\mF_3(a,b) = a + b$, and $\mF_4(a) = (a,a)$ for all~$a,b \in \mA$. For example,~$V_1$ sends over  edge~$e_3$ the alphabet symbol received via edge~$e_1$ multiplied by~2, and it forwards the received
 symbol via edge~$e_4$. 
 %The vertex~$V_3$ sums the alphabet symbols carried along the edges~$e_4$ and~$e_5$ and sends it over~$e_7$. 
 Lastly, we look at the function representing the transfer from the source to the first terminal, i.e.,
\[ \Omega[\mN, \mF, S \to T_1] : \F_3^{2} \rightarrow \F_3^{2}.\] For~$x = (x_1,x_2) \in \F_3^2$, we have~$\Omega[\mN, \mF, S \to T_1](x) \in \F_3^2$. For example, $\Omega[\mN, \mF, S \to T_1](1,2) = (2,0).$ It can be checked that~$(\F_3^2,\mF)$ is unambiguous for the Butterfly Network~$\mN$ and $\mA=\F_3$.
\end{example}

We are now ready to define 
the fundamental parameter of the network
we are interested in computing.

\begin{definition}
The~\textbf{(1-shot) capacity} of the network~$\mN$ with respect to the alphabet $\mA$ is the real number
\begin{multline} \label{c1}
    \CA(\mN, \mA) = \max\{\log_{|\mA|} |\mC| : (\mC,\mF) \mbox{ is unambiguous} \\ \mbox{for } \mN \mbox{ for some network code  $\mF$}\}.
\end{multline}
If~$\mA = \F_q$, where~$q$ is a prime power and~$\F_q$ is the finite field with~$q$ elements, and if we restrict the functions in~$\mF$ to be~$\F_q$-linear, then we call the quantity in~\eqref{c1}
the 
\textbf{(1-shot) linear capacity} of $\mN$ with respect to the alphabet~$\mA$.
We denote it by~$\CA^{\textnormal{lin}}(\mN,\mA)$.  
\end{definition}

The following result 
shows how the capacity of a network is bounded above by a combinatorial invariant; see~\cite{b2,b12}.

\begin{theorem}
\label{thm:one-shot}
We have $$\CA(\mN,\mA) \le \mu(\mN):= \min_{T \in \bfT} \textnormal{min-cut}_{\mN}(S,T),$$ 
where~$\textnormal{min-cut}_{\mN}(S,T)$ is the minimum number of edges that one needs to remove from $\mN$ in order to disconnect~$S$ and~$T$.
\end{theorem}

In \cite{b26}, it is shown that 
the bound of Theorem~\ref{thm:one-shot} can be achieved by a linear network code, provided that
the alphabet is $\F_q$ and 
$q\ge |\bfT|$.
Note that determining the minimum alphabet size for which Theorem~\ref{thm:one-shot} is met with equality is a difficult task in general. 
For example,
in \cite{b21} it is shown that 
deciding whether or not there exists a linear network code that achieves the bound of 
Theorem \ref{thm:one-shot}
is an NP-hard problem.

\begin{example}
The 1-shot capacity of the Butterfly Network of Figure \ref{fig:butterfly} is upper bounded by 2
(take the cut $\{e_1,e_2\}$ and apply Theorem~\ref{thm:one-shot}).
The bound is met with equality; see
Example \ref{ex:butterfly}. \end{example}

%%%%%%%%%%%%%%%%%%%%%%%%%%%%%%%%%%%%%%%%%%%%%

\section{A Mixed-Integer Programming Model}
\label{sec:model}

Mixed-integer programming (MIP) is a powerful framework for
modeling and solving large problems. It can be seen as linear programming (LP)~\cite{b29} with the extra constraint that at least one of the variables has to take integer values. The mathematical formulation of MIP is as follows: 
compute
\[ \max \{ \sprod{c}{\bfx} \st A \bfx \leq b \text{ and } x_i \in \Z \text{ for all $i \in I$} \},\]
where~$A$ is a rational matrix, $b$ and $c$ are rational vectors
of suitable dimensions, and $I$ is an index set.
Such problems can be solved using the celebrated branch-and-bound and
branch-and-cut methods, see~\cite{b30}.

In this section, we cast the problem of computing the capacity of networks in the framework of MIP.
We start by introducing 
a convenient notation and then the suitable variables for our model.

\begin{notation}
Let~$\mC$ and $\mF$ be an outer code and a network code for~$(\mN,\mA)$, respectively. For~$c \in \mC$, we denote the symbol vector occurring in the incoming edges of~$V \in \mV$ under the rules of the network code $\mF$ as~$c^V_{\mF}$.
\end{notation}

\begin{definition}
\label{def:MIP}
Let~$\mC$ and $\mF$ be an outer code and a network code for~$(\mN,\mA)$, respectively. 
For~$c \in \mC$, $V \in \mV$, $m \in \mA^{|\inn(V)|}$, $m' \in
\mA^{|\out(V)|}$, define
\begin{align*}
  x^{c,V}_{m'} &=
                 \begin{cases}
                   1 &\text{if } \mF_V(c^V_{\mF}) = m',\\
                   0 &\text{otherwise};
                 \end{cases}
  \\
  y^{c,V}_m &=
              \begin{cases}
                1 &\text{if } c^V_{\mF} = m,\\
                0 &\text{otherwise};
              \end{cases}
  \\
  z^V_{m,m'} &=
               \begin{cases}
                 1 &\text{if } \mF_V(m) = m',\\
                 0 &\text{otherwise}.
               \end{cases}
\end{align*}
Note that $x$ and~$y$ model the output and input at vertices, respectively, and~$z$ models the functions of the network code.
\end{definition}

To decide whether an unambiguous pair $(\mC,\mF)$ for $(\mN,\mA)$ exists, we propose to model this decision problem as an MIP and to solve it using state-of-the-art software.
Therefore, we need to specify linear constraints~$Ax + By + Cz \leq b$ that enforce the above interpretation of the variables.
More precisely, a solution~$(\bar{x}, \bar{y}, \bar{z})$ found by MIP software satisfies the interpretation if and only if~$A\bar{x} + B\bar{y} + C\bar{z}
\leq b$. 

Next, we give a system of \textit{nonlinear} constraints and provide interpretations for them.
Afterwards, we discuss how to turn this system into linear inequalities and
equations.

For~$c \in \mC,\ V \in \mV\setminus{\bfT}$,
\begin{equation}
\label{eq:findcode:uniqueOutput}
\sum_{m' \in \mA^{|\out(V)|}} x^{c,V}_{m'} =1.
\end{equation}

For~$c \in \mC,\ V \in \mV\setminus\{S\}$,
\begin{equation}
\label{eq:findcode:uniqueInput}
\sum_{m \in \mA^{|\inn(V)|}} y^{c,V}_m =1.
\end{equation}

For~$e \in \out(U) \cap \inn(V),\
    m \in \mA^{\card{\inn(V)}} $,
\begin{equation}
\label{eq:findcode:compatible}
y^{c,V}_m + \sum_{\substack{m' \in \mA^{|\out(U)|}\\ m'_e \neq m_e}} x^{c,U}_{m'} \leq 1.
\end{equation}

For~$V \in \mV \setminus (\{S\} \cup \bfT),\
    m \in \mA^{|\inn(V)|} $,
\begin{equation}
\label{eq:findcode:atMostImage}
\sum_{m' \in \mA^{|\out(V)|}} z^V_{m,m'} \leq 1.
\end{equation}

For~$c \in \mC,\
    V \in \mV\setminus (\{S\} \cup \bfT),\
    m \in \mA^{|\inn(V)|}  $,
\begin{equation}
\label{eq:findcode:atLeastImage}
\sum_{m' \in \mA^{|\out(V)|}} z^V_{m,m'} \geq y^{c,V}_m.
\end{equation}

For~$c \in \mC,\
    V \in \mV\setminus (\{S\} \cup \bfT),\
    m \in \mA^{|\out(V)|} $,
\begin{equation}
\label{eq:findcode:map}
\sum_{m \in \mA^{|\inn(V)|}} y^{c,V}_m \cdot z^V_{m,m'} = x^{c,V}_{m'}.
\end{equation}

For~$ V \in \bfT, m \in \mA^{|\inn(V)|}$,
\begin{equation}
\label{eq:findcode:noambig}
\sum_{c \in \mC} y^{c,V}_m \leq 1.
\end{equation}

Constraints~\eqref{eq:findcode:uniqueOutput}
and~\eqref{eq:findcode:uniqueInput} model that network codes completely determine the data
transmission. Constraint~\eqref{eq:findcode:compatible} makes sure that for
every edge~$e \in \out(U) \cap \inn(V)$ with~$U,V \in \mV$ and~$m \in
\mA^{|\inn(V)|}$, the output of the function~$\mF_U \in \mF$ is compatible with~$m$,
i.e., one cannot select an~$m' \in \mA^{|\out(U)|}$
that disagrees with~$m$ on the edge~$e$.
Constraints~\eqref{eq:findcode:atMostImage} and~\eqref{eq:findcode:atLeastImage} guarantee that~$\mF_V$ satisfies the definition of being a nonempty function, where~$V \in \mV\setminus (\{S\} \cup \bfT)$.
Note that an image is only defined for input strings that are actually received at~$V$.
Constraint~\eqref{eq:findcode:map} makes sure that information is correctly propagated. That is, functions of the selected network code are applied to the received inputs of the intermediate vertices of the network.
Finally, the constraint~\eqref{eq:findcode:noambig} ensures that there are no
ambiguities by enforcing that every possible input string is allowed to be used by at most one codeword\footnote{Our code is publicly available at \url{ https://github.com/christopherhojny/network_codes} (githash cfd6c62f)}.

The given model fits into the framework of MIP, except for
the nonlinear constraint~\eqref{eq:findcode:map}.
Since all variables are binary, these constraints can easily be linearized
using McCormick inequalities~\cite{b31}.
Thus, we can decide whether there exists an unambiguous pair of an outer code and a network code by using these
variables and constraints in a MIP whose objective function is constantly zero.
In other words, we ``only'' need to find vectors~$(\bar{x}, \bar{y}, \bar{z})$
satisfying all constraints.
This task can now be carried out by state-of-the-art MIP software, which
either finds a feasible vector~$(\bar{x}, \bar{y}, \bar{z})$, or proves that
no such vector exists.
In the former case, we can obtain the unambiguous pair from the interpretation
of variables; in the latter case, we conclude that no such pair
exists.

Note that MIP software does not provide a formal proof that no feasible
solution exists; this conclusion is due to the
correctness of the algorithms
used to solve a MIP problem.

\begin{remark}
The one we presented in this section is a \textit{basic} version of our model, which will appear in an extended version of this work.
For example, in its basic form the model can be difficult to solve if the size of
the code is large, because 
of symmetries: if we are given a code, we can derive an
equivalent (but for the model different) code by permuting the codewords.
We handle these symmetries by sorting the codewords lexicographically
on~$\out(S)$, which is enforced by adding so-called column inequalities to
the model~\cite{b32}.
\end{remark}

We conclude this section by briefly explaining the model with an example.

\begin{example}
Consider the Butterfly Network $\mN$ of Figure~\ref{fig:butterfly}
with $\mA=\F_3$ and 
the candidate unambiguous pair~$(\F_3^2,\mF)$, where~$\mF$ is chosen as in Example~\ref{ex:butterfly}. Consider the codeword~$c = (1,2)$. We have:
\begin{itemize}
    \item $x^{c,V_1}_{(2,1)} = 1$, $y^{c,V_1}_{\{1\}} = 1$, $z^{V_1}_{\{1\},(2,1)} =1$,
    \vspace{2px}
    \item $x^{c,V_2}_{(2,1)} = 1$, $y^{c,V_2}_{\{2\}} = 1$, $z^{V_2}_{\{2\},(2,1)} =1$.
\end{itemize}
Let~$m = (1,0) \in \alphabet^{\card{\inn(V_3)}}$ and note that~$y^{c,V_3}_{(1,0)} = 0$. Consider the edge~$e_5 \in \out(V_2) \cap \inn(V_3)$. Then
constraint~\eqref{eq:findcode:compatible} reads
$$y^{c,V_3}_m + \sum_{\substack{m' \in \mA^{|\out(V_2)|} \\ m'_{e_5} \neq 0}} x^{c,V_2}_{(m'_{e_5} ,m'_{e_6} )} \ = \sum_{\substack{m' \in \mA^{|\out(V_2)|} \\ m'_{e_5} \neq 0}} x^{c,V_2}_{(m'_{e_5} ,m'_{e_6} )} \ \leq 1.$$ 
\end{example}

%%%%%%%%%%%%%%%%%%%%%%%%%%%%%%%%%%%%%%%%%%%%%%%

\section{Reducing the Complexity}
\label{sec:coding}
We have implemented the above model using the MIP solver
Gurobi v9.1.1.
For small networks, the above model allowed us to quickly find unambiguous codes or refute their existence.
If the size of the network grows, however, solving the MIP becomes challenging.
For this reason, we discuss two simplifications based on simple but powerful results that reduce the complexity of our algorithm for some particular families of networks. 
The first trick we propose
is useful when the 
network
$\mN = (\mV,\mE, S, \bfT)$ has
\begin{equation}
\label{eq:super}
|\out(S)| > \mu(\mN).
\end{equation}
Finding codes that attain the bound of Theorem \ref{thm:one-shot} requires searching through a potentially large space.
However, we can add
a super-source~$S'$ to our network $\mN$ and connect it to the source of the previous one with~$\mu(\mN)$ edges $$e_1', \ldots, e_{\mu(\mN)}'.$$ More formally, starting from a single-source network~$\mN = (\mV,\mE, S, \bfT)$, we construct a new network 
\begin{equation}
\label{eq:newnetwork}
\mN' = (\mV \cup \{S'\},\mE \cup \{e_1', \ldots, e_{\mu(\mN)}'\}, S',\bfT).
\end{equation} 
The proof of the following result is left to the reader.

\begin{proposition}
\label{prop:supersource}
Let~$\mN = (\mV,\mE, S, \bfT)$ and~$\mN'= ({\mV \cup \{S'\}},\mE', S', \bfT)$ be as above in~\eqref{eq:newnetwork}.
Then 
$$\mu(\mN) = \mu(\mN').$$
\end{proposition}

Proposition \ref{prop:supersource} can be used for families of networks that satisfy~\eqref{eq:super} to check whether or not 
the 
bound of Theorem~\ref{thm:one-shot} is attained: the network gets slightly
larger, 
but one only needs to check whether there are network codes that form an unambiguous pair with the outer code
$$\mC = \mA ^{\mu(\mN)}$$ for~$(\mN',\mA)$ as in~\eqref{eq:newnetwork}. If the answer is yes, that means that the bound of Theorem~\ref{thm:one-shot} is met with equality for $(\mN,\mA)$. We formalize this in the following result, whose proof is omitted.

\begin{corollary}[Supersource trick]
\label{cor:same}
Let~$\mN$ and $\mN'$ be as in Proposition~\ref{prop:supersource}.
If~$(\mA ^{\mu(\mN)}, \mF)$ is unambiguous for~$(\mN',\mA)$, then~$(\mC, \mF \setminus \mF_{S'})$ is unambiguous for~$(\mN,\mA)$ where
$$\mC = \left\{\mF_S(x) \mid x\in \mA ^{\mu(\mN)}\right\}.$$
\end{corollary}

Thanks to Corollary~\ref{cor:same}, the number of possibilities for the outer code reduces from~$\binom{|\out(S)|}{\mu(\mN)}$ to one.

The second simplification
shows that
if a vertex only has a single incoming edge, then without loss of generality that vertex can use routing. The proof of the following result is omitted and will appear in the extended version of this work.
\begin{proposition}[Routing trick]
\label{prop:WLOG}
Let $V \in \mV$ be an intermediate vertex with exactly one incoming edge.
There exists an unambiguous pair $(\mC,\mF)$ for $(\mN,\mA)$ if and only if there exists an unambiguous pair $(\mC,\mF)$ for $(\mN,\mA)$ with $\mF_V$ the function that replicates any alphabet symbol.
\end{proposition}

Both strategies presented in this section can be easily incorporated into our MIP model.
The supersource trick just requires to modify the input graph, and the remaining MIP model stays the same.
The routing trick of Proposition~\ref{prop:supersource} allows us to restrict to the identity map at all vertices~$V$ with~$|\inn(V)|=1$.

We finish this section with the following remark, that also give evidence to the flexibility of our method.

\begin{remark}
\label{rem:trick}
Let~$V \in \mV \setminus (\{S\} \cup \bfT)$ with~$|\inn(V)|=1$. The routing trick of Proposition~\ref{prop:WLOG} can easily be incorporated into our model by fixing all variables~$z^V_{m,m'}$ to be~0 if~$m'_e \neq m$ for all~$e \in \out(V).$    
\end{remark}

\section{Evidence and Concrete Applications}
\label{sec:examples}

In this section we apply our approach in combination with the 
simplifications presented in Section~\ref{sec:coding}
to compute the exact capacity of 
some ``nasty'' networks over small alphabets.
This improves and refines known results about these networks.

\begin{figure}[htbp]
\centerline{\includegraphics[width=.3\textwidth]{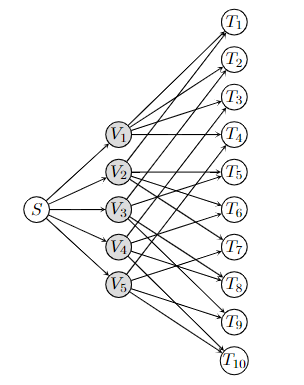}}
\caption{The $\binom{5}{2}$ combination network.}
\label{fig:combination}
\end{figure}

We start with the \textit{combination networks} defined in \cite{b23}, which is a family of networks that satisfy~\eqref{eq:super}. We will focus on the $\binom{5}{2}$ combination network to simplify the exposition. The latter is depicted in Figure \ref{fig:combination}.
We start by summarizing what is known already.

\begin{theorem}[see~\cite{b11}]
\label{comb_known}
Let~$\mN$ be the network of Figure~\ref{fig:combination}. We have~$\CA(\mN,\mA)=2$ if and only if~$|\mA| \notin \{2,3,6\}.$
\end{theorem}

We start by reporting that ``$\CA(\mN,\mA)<2$ if~$|\mA| \in \{2,3,6\}$''
can be easily verified using our model in combination with Corollary~\ref{cor:same}. 
Next, we are interested in computing the exact capacity when 
$|\mA| \in \{2,3,6\}$.
We use our model restricted to routing, thanks to Proposition~\ref{prop:WLOG}. We summarize our findings in the following theorem, which extends Theorem~\ref{comb_known}.

\begin{theorem}
\label{thm:comb}
Let~$\mN$ be the network of Figure~\ref{fig:combination}. We have
$$\CA(\mN,\mA)=
\begin{cases}
      1  & \text{if} \ \ |\mA| = 2, \\
      \log_3 6  & \text{if} \ \ |\mA| = 3, \\
      \log_6 X & \text{if} \ \ |\mA| = 6,\\
      2  & \text{otherwise.}
\end{cases}$$ for some integer $X$ with $29 \le X \le 35$.  
\end{theorem}

By repeatedly applying
Proposition~\ref{prop:WLOG} we get the linear capacity of the 
$\binom{5}{2}$ combination network as a corollary.

\begin{corollary}
\label{cor:comb}
Let~$\mN$ be the network of Figure~\ref{fig:combination} and~$\mA =\F_q$, where~$q$ is a prime power. We have
$$\CA^{\textnormal{lin}}(\mN,\mA)=
\begin{cases}
      1  & \text{if} \ \ q = 2, \\
      \log_3 6  & \text{if} \ \ q = 3, \\
      2  & \text{otherwise.}
\end{cases}$$
\end{corollary}

Corollary~\ref{cor:comb} shows that the bound~$q \ge |\bfT|$ of~\cite{b26} may be far from being 
tight. Indeed, we know there is a linear network coding solution for the~$\binom{5}{2}$ combination network for~$q \in \{4,5,7,8,9\}$, while~$|\bfT| = 10$.

\begin{figure}[htbp]
\centerline{\includegraphics[width=.4\textwidth]{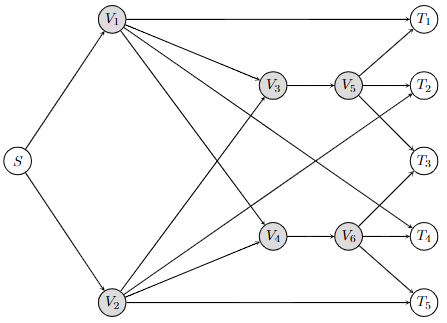}}
\caption{A network from \cite{b11}.}
\label{fig:riis}
\end{figure}

For the last part of the paper, we focus on the network of Figure~\ref{fig:riis}, studied in~\cite{b11}.

The following result was established using the theory of Latin squares (see \cite{wallis2016introduction} for an exposition on Latin squares).

\begin{theorem}[see~\cite{b11}]
\label{thm:b11}
Let~$\mN$ be the network of
 Figure~\ref{fig:riis}. We have~$\CA(\mN,\mA) < 2$ if~$|\mA| \in \{2,6\}$ and otherwise~$\CA(\mN,\mA)=2$.
\end{theorem}

We combine our model with Proposition~\ref{prop:WLOG} and Theorem~\ref{thm:b11}, obtaining the following result.

\begin{theorem}
\label{thm:riis}
Let~$\mN$ be the network of Figure~\ref{fig:riis}. We have
$$\CA(\mN,\mA)=
\begin{cases}
      1  & \text{if} \ \ |\mA| = 2, \\
      \log_6 34  & \text{if} \ \ |\mA| = 6, \\
      2  & \text{otherwise.}
\end{cases}$$  
\end{theorem}

Since the network of Figure~\ref{fig:riis} has 
five terminals, we already know from~\cite{b26} that linear coding solution exists for
$\mA = \F_q$
and
$q \ge 5$. 
By checking the output of our model,
we obtain the following result (where the proof is inspired by the output of the model).

\begin{theorem}
Let~$\mN$ be the network of
Figure~\ref{fig:riis} and~$\mA =\F_q$, where~$q$ is a prime power. We have
$$\CA^{\textnormal{lin}}(\mN,\mA)=
\begin{cases}
      1  & \text{if} \ \ q = 2, \\
      2  & \text{otherwise.}
\end{cases}$$
\end{theorem}
\begin{IEEEproof}
Proposition~\ref{prop:WLOG} guarantees~$\F_q$-linear functions except possibly for the functions used by~$V_3$ and~$V_4$. The capacity for~$|\mA| = 2$ is achieved by routing with any outer code of size~$2$. The capacity for~$|\mA| = 3$ is achieved via the pair~$(\F_3^2,\mF)$, where:
$$\mF_{V_3}(a,b) = a+2b, \quad \mF_{V_4}(c,d) = c+d$$ for all $a,b,c,d$. The capacity for~$|\mA| = 4$ is achieved via the pair~$(\F_4^2,\mF)$, where:
$$\mF_{V_3}(a,b) = \alpha a+b, \quad \mF_{V_4}(c,d) = (\alpha+1)c+d$$ for~$\F_4 = \F_2[\alpha]/(\alpha^2 + \alpha +1)$,
$a \in \out(V_1) \cap \inn(V_3),\ b \in \out(V_2) \cap \inn(V_3),\ c \in \out(V_1) \cap \inn(V_4),\ d \in \out(V_2) \cap \inn(V_4)$.
This concludes the proof.
\end{IEEEproof}

Theorems~\ref{thm:comb} and~\ref{thm:riis} also show that the existence of an optimal solution for a given alphabet size does not guarantee the existence of optimal solutions for larger alphabets.

\section{Conclusions and Future Research Program}
\label{sec:conclusion}

In this paper, we have proposed a combinatorial optimization framework to compute the capacity of 
coded networks over small alphabets.
Note that this problem is still wide open, especially for the case where the alphabet size is not a prime power; see e.g.~\cite{b24,b25}.
The framework uses mixed-integer programming combined with some ``theoretical'' results to reduce the running time. As a first application, we extend and refine various results on the capacity of certain networks.

This paper is the stepping stone of an ambitious program 
aimed at computing the capacity of simple adversarial networks over small alphabets. This task is particularly relevant for emerging research directions about networks with restricted adversaries; see~\cite{net_dec}. In that context, five families of small networks form the building blocks of the theory. Evaluating the capacity of those network families would give important insight on the capacity of arbitrarily large networks with restricted adversaries; see in particular~\cite[Section~7]{net_dec}. 

The natural next step is to extend our optimization approach to model adversarial networks as well.
We will then evaluate the capacity of the five families introduced in~\cite{net_dec} and use these results to study larger networks. These results will appear in an extended version of this work.

\newpage

\bibliography{bbb}
\bibliographystyle{ieeetr}

\end{document}